\documentclass[12pt]{article}
\usepackage{amssymb,diagrams}
\newcommand{\be}{\begin{eqnarray}}
\newcommand{\ee}{\end{eqnarray}}

\newcommand{\pa}{\partial}
\renewcommand{\d}{{\rm d}}
 
\newcommand{\dl}{{\delta}}
\newcommand{\D}{{\rm D}}

\title{\bf Bicomplexes, Integrable Models, and Noncommutative Geometry}
\date{  }
\author{Aristophanes Dimakis \\ Department of Mathematics, University of the Aegean \\
        GR-83200 Karlovasi, Samos, Greece 
        \\[2ex]
        Folkert M\"uller-Hoissen \\ Max-Planck-Institut f\"ur Str\"omungsforschung \\
        Bunsenstrasse 10, D-37073 G\"ottingen, Germany }

\begin{document}
\renewcommand{\theequation} {\arabic{section}.\arabic{equation}}

\maketitle

\begin{abstract}
We discuss a relation between bicomplexes and integrable models, 
and consider corresponding noncommutative (Moyal) deformations. As an 
example, a noncommutative version of a Toda field theory is presented.
\end{abstract}

\section{Introduction}
\setcounter{equation}{0}
Soliton equations and integrable models are known to possess a
vanishing curvature formulation depending on a parameter, say $\lambda$ 
(cf \cite{Fadd+Takh87}, for example). This geometric formulation of 
integrable models is easily extended \cite{DMH96,DMH00} to generalized geometries, 
in particular in the sense of noncommutative geometry where, on a basic level, 
the algebra of differential forms (over the algebra of smooth functions) 
on a manifold is generalized to a differential calculus over an 
associative (and not necessarily commutative) algebra $\cal A$. 
\vskip.1cm

A bicomplex associated with an integrable model is a special case of a 
zero curvature formulation. More precisely, let 
$ {\cal M} = \bigoplus_{r \geq 0} {\cal M}^r $
be an ${\mathbb{N}}_0$-graded linear space (over ${\mathbb{R}}$ or 
${\mathbb{C}}$) and 
$\d , \delta \, : \, {\cal M}^r \rightarrow {\cal M}^{r+1}$ 
two linear maps satisfying\footnote{In terms of 
$\d_\lambda = \delta - \lambda \, \d$ with a constant $\lambda$,
the three bicomplex equations are combined into the single zero curvature
condition $\d_\lambda^2 = 0$ (for all $\lambda$).} 
\be
   \d^2 = 0 \, , \quad 
   \delta^2 = 0 \, , \quad   
   \d \, \delta + \delta \, \d = 0
          \label{bic-eqs}
\ee 
(typically as a consequence of certain field equations). 
Then $({\cal M},\d,\delta)$ is called a {\em bicomplex}.
Special examples are bi-differential calculi \cite{DMH00}. However, we do not 
need $\d$ and $\delta$ to be graded derivations (into some bimodule), 
i.e., they do not have to satisfy the Leibniz rule. 
\vskip.1cm

Given a bicomplex, there is an iterative construction of ``generalized conserved 
densities" in the sense of $\delta$-closed elements of the bicomplex (see section 2).
In some examples they reproduce directly the conserved quantities 
of an integrable model. In other examples, the relation is less direct.
Anyway, the existence of such a chain of $\delta$-closed elements is 
clearly a distinguished feature of the model with which the bicomplex is 
associated. 
\vskip.1cm

Noncommutative examples are in particular obtained by starting 
with a classical integrable model, deforming an associated bicomplex by 
replacing the ordinary product of functions with the Moyal $\ast$-product 
and thus arriving at a noncommutative model.
As an example, a noncommutative extension of a Toda field 
theory is considered in section 3. 
Field theory on noncommutative spaces has gained more and more interest 
during the last years. A major impulse came from the discovery that a 
noncommutative gauge field theory arises in a certain limit of string, 
D-brane and M theory (see \cite{Seib+Witt99} and the references cited there). 
We also refer to \cite{nc-intmod} for some work on Moyal deformations of 
integrable models.

\section{The bicomplex linear equation}
\setcounter{equation}{0}
Let us assume that, for some $s \in {\mathbb{N}}$, there 
is a (nonvanishing) $\chi^{(0)} \in {\cal M}^{s-1}$ with $\d J^{(0)} =0$ 
where $J^{(0)} = \delta \chi^{(0)}$.  
Defining $J^{(1)} = \d \chi^{(0)}$, 
we have $\delta J^{(1)} = - \d \delta \chi^{(0)} = 0$ using (\ref{bic-eqs}).
If the $\delta$-closed element $J^{(1)}$ is $\delta$-exact, this implies 
$J^{(1)} = \delta \chi^{(1)}$ with some $\chi^{(1)} \in {\cal M}^{s-1}$. Next 
we define $J^{(2)} = \d \chi^{(1)}$. Then 
$\delta J^{(2)} = - \d \delta \chi^{(1)} = - \d J^{(1)} = - \d^2 \chi^{(0)} = 0$.
If the $\delta$-closed element $J^{(2)}$ is $\delta$-exact, then 
$J^{(2)} = \delta \chi^{(2)}$ with some $\chi^{(2)} \in {\cal M}^{s-1}$. 
This can be iterated further and leads to a possibly infinite 
chain (see the figure below) of elements $J^{(m)}$ of ${\cal M}^s$ and 
$\chi^{(m)} \in {\cal M}^{s-1}$ satisfying
\be
         J^{(m+1)} = \d \chi^{(m)} = \delta \chi^{(m+1)}   \; .
             \label{Jm-chim}
\ee
More precisely, the above iteration continues from the $m$th to the
$(m+1)$th level as long as $\delta J^{(m)} = 0$ 
implies $J^{(m)} = \delta \chi^{(m)}$ with an element 
$\chi^{(m)} \in {\cal M}^{s-1}$. Of course, there is no obstruction to the 
iteration if $H^s_\delta ({\cal M})$ is trivial, 
i.e., when all $\delta$-closed elements of ${\cal M}^s$ are $\delta$-exact.
But in general the latter condition is too strong, though
in several examples it can indeed be easily verified \cite{DMH00}.
Introducing 
\be
   \chi = \sum_{m \geq 0} \lambda^m \, \chi^{(m)}
\ee
with a parameter $\lambda$, the essential ingredients of the above iteration 
procedure are summarized in the {\em linear equation} associated with the 
bicomplex:
\be
    \delta (\chi - \chi^{(0)}) = \lambda \, \d \, \chi  \; .
                   \label{bc-linear}
\ee
\small
\diagramstyle[PostScript=dvips]
\begin{diagram}[notextflow]
 &         &\chi^{(0)}&         &       &         &\chi^{(1)}&         &       &         &\chi^{(2)}&&& \\
 &\ldTo^\dl&          &\rdTo^\d &       &\ldTo^\dl&          &\rdTo^\d &       &\ldTo^\dl&          &\rdTo^\d&& \\
J^{(0)}&         &          &         &J^{(1)}&         &          &         &J^{(2)}&         &          & &J^{(3)}&\cdots\\
 &\rdTo^\d &          &\ldTo^\dl&       &\rdTo^\d &          &\ldTo^\dl&       &\rdTo^\d &          &\ldTo^\dl&&  \\
 &         & 0        &         &       &         & 0        &         &       &         & 0        &&&
\end{diagram}
\normalsize

\vskip.1cm

Given a bicomplex, we may start with the linear 
equation (\ref{bc-linear}). Let us assume that it admits a (non-trivial)
solution $\chi$ as a (formal) power series in $\lambda$.
The linear equation then leads to $\delta \chi^{(m)} = \d \chi^{(m-1)}$. 
As a consequence, the $J^{(m+1)} = \d \chi^{(m)}$ are $\delta$-exact. 
Therefore, even if the cohomology $H^s_\delta ({\cal M})$ is {\em not} trivial,
the solvability of the linear equation ensures that the $\delta$-closed $J^{(m)}$
appearing in the iteration are $\delta$-exact.
\vskip.1cm

In all the examples which we presented in \cite{DMH00,DMH99}, the 
bicomplex space can be chosen as ${\cal M} = {\cal M}^0 \otimes \Lambda$ where 
$\Lambda = \bigoplus_{r=0}^n \Lambda^r$
is the exterior algebra of an $n$-dimensional vector space with a basis $\xi^r$,
$r=1, \ldots, n$, of $\Lambda^1$. It is then sufficient
to define the bicomplex maps $\d$ and $\delta$ on ${\cal M}^0$ since via 
\be
 \d (\sum_{i_1,\ldots,i_r =1}^n \phi_{i_1 \ldots i_r} \, \xi^{i_1} \cdots \xi^{i_r} ) 
 = \sum_{i_1,\ldots,i_r =1}^n ( \d \phi_{i_1 \ldots i_r} ) \, \xi^{i_1} \cdots \xi^{i_r}
\ee
(and correspondingly for $\delta$) they extend as linear maps to the whole of ${\cal M}$.

\section{Noncommutative deformation of a Toda model}
\setcounter{equation}{0}
The $\ast$-product on the space $\cal F$ of smooth functions of 
two coordinates $x$ and $t$ is given by
\be
    f \ast h = m \circ e^{\theta P/2} (f \otimes h) 
             = f h + {\theta \over 2} \, \{ f , h \} + {\cal O}(\theta^2)
\ee
where $\theta$ is a parameter, $m(f \otimes h) = f h$ and
$P = \pa_t \otimes \pa_x - \pa_x \otimes \pa_t$. Furthermore, 
$\{ \, , \, \}$ is the Poisson bracket, i.e., 
$\{ f , h \} = (\pa_t f) \, \pa_x h - (\pa_x f) \, \pa_t h$.
 For the calculations below it 
is helpful to notice that partial derivatives are derivations of 
the algebra $({\cal F}, \ast)$. 
\vskip.1cm

A bicomplex associated with an integrable model can be deformed by 
replacing the ordinary product of functions with the noncommutative $\ast$-product. 
This then induces a deformation of the integrable model with very special properties 
since the iterative construction of generalized conservation laws still works.
As a specific example, we construct a noncommutative extension of the 
Toda field theory on an open finite one-dimensional lattice. Other examples 
can be obtained in the same way.
\vskip.1cm

Let us start from the trivial bicomplex which is determined by
\be
   \delta \phi = (\pa_t - \pa_x) \phi \, \xi^1 + (S - I) \phi \, \xi^2 
         \, , \quad
   \d \phi = -S^T \phi \, \xi^1 + (\pa_t + \pa_x) \phi \, \xi^2   
\ee 
where $\phi$ is a vector with $n$ components (which are functions) 
and $S^T$ the transpose of
\be
    S = \sum_{i=1}^{n-1} E_{i,i+1} \, , \qquad 
   (E_{i,j})^k{}_l = \delta^k_i \, \delta_{j,l} \; .
\ee

Let $G$ be an $n \times n$ matrix of functions which is invertible in the sense
$G^{-1} \ast G = I$ where $I$ is the $n \times n$ unit matrix. 
Now we introduce a ``dressing" for $\d$:
\be
    \D \phi = G^{-1} \ast \d (G \ast \phi)
            = -(L \ast \phi) \, \xi^1 + (\pa_t + \pa_x + M \ast ) \phi \, \xi^2  
\ee
where
\be
     L = G^{-1} \ast S^T \ast G \, , \quad
     M = G^{-1} \ast (G_t + G_x)   \; .
\ee
Note that $\D^2 \phi = G^{-1} \ast \d^2 (G \ast \phi) = 0$. 
The only nontrivial bicomplex equation is $\delta \D + \D \delta = 0$ 
which reads
\be
     M_t - M_x = L \ast S - S \ast L  \; .
	       \label{nc-bic}
\ee
Hence, if this equation holds, then $({\cal F}^n \otimes \Lambda, \D, \delta)$ 
is a bicomplex. Let us now choose
\be
    G = \sum_{i=1}^n G_i \, E_{ii}
\ee
with functions $G_i$ for which the invertibility assumption requires 
$G_i^{-1} \ast G_i = 1$. Then
\be
    L = \sum_{i=1}^{n-1} G_{i+1}^{-1} \ast G_i \, E_{i+1,i} \, , \quad
    M = \sum_{i=1}^n M_i \, E_{ii} \, , \quad
    M_i = G_i^{-1} \ast (\pa_t + \pa_x) G_i
        \; .
\ee
Writing
\be
    G_i = e^{q_i} (1 + \theta \, \tilde{q}_i) + {\cal O}(\theta^2)
\ee
we have $G_i^{-1} = e^{-q_i} (1 - \theta \, \tilde{q}_i) + {\cal O}(\theta^2)$ 
and it follows from (\ref{nc-bic}) that the functions $q_i$ have to solve 
the Toda field theory equations
\be
   (\pa_t^2 - \pa_x^2) q_i &=&  e^{q_{i-1} - q_i} - e^{q_i - q_{i+1}}  \qquad
                                i = 2, \ldots, n-1 \nonumber  \\
   (\pa_t^2 - \pa_x^2) q_1 &=&  - e^{q_1 - q_2} \, , \quad 
   (\pa_t^2 - \pa_x^2) q_n  =  e^{q_{n-1} - q_n}   \; .
            \label{Toda}  
\ee 
Furthermore, the functions $\tilde{q}_i$ are subject to the following 
linear equations,
\be
   (\pa_t^2 - \pa_x^2) \tilde{q}_1 &=& \{ \pa_t q_1 , \pa_x q_1 \}
	    - e^{q_1 - q_2} (\tilde{q}_1 - \tilde{q}_2) 
                                      \nonumber \\
   (\pa_t^2 - \pa_x^2) \tilde{q}_i &=& \{ \pa_t q_i , \pa_x q_i \}
    + e^{q_{i-1} - q_i} (\tilde{q}_{i-1} - \tilde{q}_i) 
	- e^{q_i - q_{i+1}} (\tilde{q}_i - \tilde{q}_{i+1}) 
                                      \nonumber \\	
	(\pa_t^2 - \pa_x^2) \tilde{q}_n &=& \{ \pa_t q_n , \pa_x q_n \}
	    + e^{q_{n-1} - q_n} (\tilde{q}_{n-1} - \tilde{q}_n)  \; .
\ee
\vskip.1cm

A 1-form $J = P \, \xi^1 + R \, \xi^2$ is $\delta$-closed iff 
$(\pa_t - \pa_x) R = (S-I) \, P$. For $J = \lambda \, \D \chi$ 
(cf (\ref{Jm-chim})) we have 
$P = - \lambda \, L \ast \chi$ and $R = \lambda \, (\pa_t + \pa_x + M \ast) \chi$
and thus
\be
    \pa_t [\lambda \, ( \pa_t + M_i \ast) \chi_i ]
  = \pa_x [\lambda \, ( \pa_x + M_i \ast) \chi_i ] 
    + P_{i+1} - P_i
\ee
($i = 1, \ldots , n$) where we have to set $P_{n+1} =0$. 
Using $P_1 = 0$, we find that 
\be
    Q = \lambda \int dx \, \sum_{i=1}^n \left( \pa_t \chi_i 
        + G_i^{-1} \ast [(\pa_t + \pa_x) G_i] \ast \chi_i \right)
\ee
is conserved, i.e., $dQ/dt = 0$, provided that the expressions 
$\pa_x \chi_i + M_i \ast \chi_i$ vanish at $x= \pm \infty$. 
In order to further evaluate this expression, we have to explore  
the linear system associated with the bicomplex. Choosing 
$\chi^{(0)} = \sum_{i=1}^n e_i$, where $e_i$ is the vector with components 
$(e_i)_j = \delta_{ij}$, we have 
$\delta \chi^{(0)} = - e_n \, \xi^2$ and $\D \delta \chi^{(0)} = 0$.
Now we find
\be
    Q^{(1)} 
 &=& \int dx \, \sum_{i=1}^n M_i 
  = \int dx \, \sum_{i=1}^n G_i^{-1} \ast (\pa_t + \pa_x) G_i \nonumber \\
 &=& \int dx \, \sum_{i=1}^n \pa_t q_i 
     + \theta \int dx \, \sum_{i=1}^n \left( (\pa_t + \pa_x) \tilde{q}_i 
     + {1 \over 2} \, \{ (\pa_t + \pa_x) q_i , q_i \} \right) 
     + {\cal O}(\theta^2)          \qquad
\ee
where we assumed that the $q_i$ vanish at $x = \pm \infty$.
The linear system $\delta ( \chi - \chi^{(0)} ) = \lambda \, \D \chi$ reads
\be
   (\pa_t - \pa_x) \chi_1 = 0 \, , \quad
   (\pa_t - \pa_x) \chi_i = - \lambda \, G_i^{-1} \ast G_{i-1} \ast \chi_{i-1}
            \qquad i = 2, \ldots, n   
\ee
and
\be
   \chi_{i+1} - \chi_i &=& \lambda \, ( \pa_t + \pa_x + M_i \ast ) \chi_i
              \qquad i=1, \ldots, n-1  
			  \label{ncToda-chi-eq}   \\
   \chi_n &=& 1 - \lambda \, ( \pa_t + \pa_x + M_n \ast ) \chi_n   \; .
\ee
Using $\chi^{(0)}_i = 1$, (\ref{ncToda-chi-eq}) yields
$\chi^{(1)}_{i+1} - \chi^{(1)}_i = M_i$ and thus\footnote{A ``constant of 
integration" can be added on the rhs. But this would simply lead to an 
additional term proportional to $Q^{(1)}$ in (\ref{Q2}).}  
$\chi^{(1)}_i = - \sum_{k=i}^n M_k$. 
After some manipulations and using (\ref{Toda}), we obtain
\be
      Q^{(2)} 
  &=& - \int dx \sum_{i=1}^n \sum_{k=i}^n ( \pa_t M_k + M_i \ast M_k )
                                 \nonumber \\
  &=& - \int dx \left( \sum_{i=1}^{n-1} e^{q_i - q_{i+1}} 
      + {1 \over 2} \left( \sum_{i=1}^n (\pa_t+\pa_x) q_i \right)^2
      - {1 \over 2} \sum_{i=1}^n [(\pa_t+\pa_x) q_i]^2 \right) 
                        \nonumber \\
  & &  + {\cal O}(\theta)         \label{Q2}
\ee
where to first order in $\theta$ already a rather complicated expression emerges.
At 0th order in $\theta$ one recovers the known conserved 
charges of the Toda theory.
\vskip.2cm

Infinite-dimensional integrable models possess an infinite set of conserved 
currents. In contrast to previous approaches to deformations of integrable models 
(see \cite{nc-intmod}, for example), our approach guarantees, via deformation 
of the bicomplex associated with an integrable model, that this infinite tower 
of conservation laws survives the deformation.\footnote{Deforming a Hamiltonian 
system which is (Liouville) integrable so that the conserved charges 
are in involution with respect to a symplectic structure, the question arises whether 
there is a corresponding deformation of the symplectic and Hamiltonian structure 
such that the deformation preserves the involution property.}

\end{document}